\documentclass[12pt,a4paper]{article}
\usepackage[utf8]{inputenc}
\usepackage{amsmath}
\usepackage{amsfonts}
\usepackage{amssymb}
\usepackage{amsbsy}
\usepackage{graphicx}
\usepackage{tabularx}
\usepackage[sharp]{easylist}

\newtheorem{proposition}{Proposition}

\newtheorem{corollary}{Corollary}
\newtheorem{definition}{Definition}

\begin{document}
\title{A Novel Approach of Pseudorandomly sorted list-based Steganography}
\author{Ren\'{e} Ndoundam\footnote{Corresponding author: ndoundam@yahoo.com} , St\'{e}phane Gael R. Ekodeck\\
{\small University of Yaounde I, LIRIMA, Team GRIMCAPE, P.o.Box 812 Yaounde, Cameroon} \\
{\small CETIC, Yaounde, Cameroon} \\
{\small IRD, UMI 209, UMMISCO, IRD France Nord, F-93143, Bondy, France} \\
{\small Sorbonne Unversit\'es, Univ. Paris 06, UMI 209, UMMISCO, F-75005, Paris, France} \\
{\small E.mail: ndoundam@yahoo.com, ekodeckstephane@gmail.com} 
 }
\date{}
\maketitle {}
\begin{abstract}
We propose a new model of steganography based on a list of pseudo-randomly sorted sequences of characters. Given a list $L$ of $m$ columns containing $n$ distinct strings each, with low or no semantic relationship between columns taken two by two,  and a secret message $s \in \{0,1\}^*$, our model embeds $s$ in $L$ block by block, by generating, for each column of $L$, a permutation number and by reordering strings contained in it according to that number. Where, letting $l$ be average bit length of a string, the embedding capacity is given by $[(m-1)*log_2(n!-1)/n*l]$. We've shown that optimal efficiency of the method can be obtained with the condition that $(n >> l)$. The results which has been obtained by experiments, show that our model performs a better hiding process than some of the important existing methods, in terms of hiding capacity. 
\end{abstract}
{\bf Keywords:} Steganography, pseudorandom sort, permutation, list, strings, embedding capacity.

\section{Introduction}
Steganography \cite{PW 09} is derived from a work by Johannes Trithemus (1462-1516) entitled \textit{Steganographia} and comes from the Greek name \textit{steganos} (hidden or secret) and \textit{graphy} (writing or drawing) and literally means \textit{hidden writing} \cite{PD 08, PR 03}. It is an ancient art of hiding information, which it's goal consists in hiding a secret message in a public media (video, text, sound, image, etc.) acting as a cover, in a way that sent through a non-secure communication channel, only the sender and the receiver are able to understand it, and anyone else cannot distinguish the existence of an hidden message.\\

Steganography has been many times used throughout history such that several schemes throughout it were developed. Its scientific study in the open literature began in 1983 when Simmons \cite{GJS 83} stated the problem in terms of communication in a prison. In his formulation, two inmates, Alice and Bob, are trying to hatch an escape plan. The only way they can communicate with each other is through
a public channel, which is carefully monitored by the warden of the prison, Ward. If Ward detects any encrypted messages or codes, he will throw both Alice and Bob into solitary confinement. The problem of steganography is then: how can Alice and Bob cook up an escape plan by communicating over
the public channel in such a way that Ward does not suspect that anything “unusual" is going on\cite{HLA 02}?\\

Most of the work in steganography has been done on images, video clips, music, sounds and texts, taken individually. We have oriented towards text-based steganography as in our study we have found out that it's difficult to use text files as cover media, despite the fact that, with the multiplication of file transfers on networks, they are highly used. This is due to fact that, by using text files as cover media, hiding techniques face problems of \cite{LHZYC 00, GKDS 12}:
\begin{itemize}
\item lack of redundancy whereas lot of redundancy is present in image or sound files, leading to a high use of those files in steganography;
\item low embedding capacity of secret data in text files;
\item imperceptibility of the modifications done on the cover file not well managed;
\item necessity of a cover file with high size to hide only few information.
\end{itemize}

 In this paper, we present a hiding technique using list of strings of any type (bit, character, etc.) as cover, such that ordered in a pseudo-random way by application of a permutation, and sent through a non-secure communication channel, only the sender and the receiver are able to retrieve the hidden message.\\

In the sequel, in Section 2 we present some related hiding techniques followed by some permutation methods in Section 3. After that, our contribution in Section 4 is described, followed by the definition of a list in Section 5 that helps us to exhibit our hiding technique in Section 6. We end this paper by presenting experimental results in Section 7 and a conclusion in Section 8. 

\section{Related Work}

Throughout history, many people worked on how to develop hiding techniques using text as cover media. In order to get a clear view of the landscape, in the literature we've found the following studies.\\

Wayner \cite{PW 92, PW 02} introduced the mimic functions where, he used the inverse of Huffman Code by inputting a data stream of randomly distributed bits. His purpose was to produce a text that fits the statistical profile of a particular normal text. Thus, the generated text by mimic functions is resilient against statistical attacks. The output of a regular mimic functions is gibberish. Accordingly, this makes the text extremely suspicious \cite{AD 09}. \\

In 1995, text data hiding program called TEXTO is exhibited by Maher \cite{MK 95}, which was designed to transform unencoded or PGP ASCII-armored ASCII data into English sentences. It is practical for exchanging binary data, especially encrypted data. Here, the secret data is replaced by English words.\\

Chapman and Davida \cite{CD 97, CD 01, CD 02} introduced a steganographic scheme that consists of two functions called NICETEXT and SCRAMBLE. NICETEXT transforms a cipher text into a text that looks like natural language.\\
There are synonyms-based approach which attracted the attention of many researchers like Winstein \cite{KW 99}, Nakagawa et al. \cite{NSMKMM 01} and Murphy et al. \cite{MV 07}. In synonym-based approach, the cover text may look legitimate from a linguistics point of view given the adequate accuracy of the chosen synonyms. But reusing the same piece of text to hide a message can raise suspicion \cite{AD 09}. Sun et al. \cite{SLH 04} proposed a scheme that uses the left and right components of Chinese characters. The proposed scheme is called L-R scheme. In L-R scheme, the mathematical expression of all Chinese characters is introduced into the text data hiding strategy. It chooses those characters with left and right components as candidates to hide the secret information.\\

In order to increase the hiding capacity of L-R scheme of Sun et al., Wang et al. \cite{WCLL 09} revised it by adding the up and down structure of Chinese characters as an extra candidate set. Besides, a reversible function to Sun et al.’s L-R scheme has been added to make it possible for receivers to obtain the original cover text and use it repeatedly for later transmission of secrets after the initial hidden secrets have been extracted \cite{WCLL 09}.
Since communications via chat room become more popular in people’s lives.\\
Wang and Chang proposed another new text steganography method. The proposed method embeds secret information into emotional icons (also called emoticons) in chat rooms over the Internet. In this method, firstly the sender’s emoticon table should be unanimous with the receiver’s emoticon table. Next, the sender and the receiver classify those emoticons in the emoticon table into several sets according to their meaning (like cry, smile laugh) and every emoticon belongs to one set. The order number of an emoticon, counting from 0, in its set is the secret bits that will be embedded. Thus, the proposed steganographic scheme uses a secret key to control the order of emoticons in each constructed set. Only the sender and the receiver keep this key. The embedding capacity has also been improved due to the tremendous numbers of emoticons used in many kinds of chat rooms \cite{WCLL 09}.\\

Stutsman et al. \cite{SAGG 06} introduced a new approach that is called translation-based steganographic scheme. This scheme hides a message in the errors (noise), which are naturally encountered in a machine translation (MT). The secret message is embedded by performing a substitution procedure on the translated text using translation variations of multiple MT systems \cite{AD 09}. Another noise-based approach was proposed by Topkara et al. in \cite{TTA 07}. Here, typos and ungrammatical abbreviations in a text, for example, emails, blogs, forums, are employed for hiding data. However,
this approach is sensitive to the amount of noise (errors) that occurs in a human writing \cite{AD 09}.\\

In 2009, Desoky presented the development of List-Based Steganography Methodology (Listega), which conceals data in textual list of itemized data. The high demand for textual list of itemized data by a wide variety of people allows the communicating parties to establish a covert channel to transmit hidden messages (listcover) rendering textual list of items an attractive steganographic carrier. Listega neither hides data in a noise (errors) nor produces noise. Instead, it camouflages data in legitimate list of items by manipulating, mainly the itemized data e.g., list of books, movie DVD’s, music CD’s, auto-parts) in order to embed data without generating any suspicious pattern \cite{AD 09}.\\

Por et al. \cite{PWC 12} proposed a data hiding method based on space character manipulation called UniSpaCh. UniSpaCh is proposed to embed information in Microsoft Word document using Unicode space characters. In addition, white spaces are considered to encode payload because they appear throughout the document (i.e., available in large number), and the manipulation of white spaces has insignificant effect to the visual appearance of document. UniSpaCh embeds payload into inter-sentence, inter-word, end-of-line and inter-paragraph spacings by introducing Unicode space characters \cite{PWC 12}.\\

In 2011, Kumar et al. \cite{KCS 11} in their paper proposed an email based high capacity text steganography method using combinatorial compression. The method makes use of forward email platform to hide the secret data in email addresses. They used the combination of BWT + MTF + LZW coding algorithm to increase the hiding capacity, as it is proved that this combination increases the compression ratio. To further increase the capacity, the numbers of characters of email id are also used to refer the secret data bits. Furthermore, the method adds some random characters just before the ‘@' symbol of email ids to increase the randomness.\\

Satir et al. \cite{SI 12} considered in their study the improvement of capacity and security issues of text steganography, by proposing a novel approach that employs data compression. They choose LZW data compression algorithm as it's frequently used in the literature and has a significant compression ratio. Their method constructs – uses stego keys and employs Combinatorics-based(use of Latin Square) coding in order to increase security. Secret information has been hidden in the chosen text from the previously constructed text base that consists of naturally generated texts. Email has been chosen as communication channel between the two parties, so the stego cover has been arranged as a forward mail platform.\\

In \cite{HH 12}, H. Hioki introduced data embedding methods, called distortion-free (or distortionless) steganographic  methods, that are not based on modification of the contents
 of cover objects. Which are useful methods when the quality of stego object matters crucially. 
 
\begin{itemize}
\item Permutation steganography: here, a secret message is embedded as a number corresponding to a permutation that is represented by a tuple of cover elements. Embedding is performed
 by shuffling cover elements such that the content of a cover object is preserved if the rearrangement of its elements does not affect the content;
\item Metadata steganography: this method does not modify data in cover objects. Instead, it adjusts the metadata of the files, to represent a secret message.
Due to the low level of embedding capacity of metadata, the method simultaneously uses for embedding, the metadata of files contained in a directory tree;
\item Cover generation methods: in this method, a cover object is generated so that it becomes a stego object that encodes a message as it is. Hioki described a text-based and image-based
 methods. The text-based method is based on a customized context-free grammar. In this method, a message is encoded into sentences using production rules of grammars. In image-based methods, 
 an image is mapped to a bit string, and a message is encoded by a sequence of images to be saved as an image gallery or image mosaic.
\end{itemize}
\section{Permutations}
\subsection{Permutation}
A permutation, also called an \textit{arrangement number} or \textit{order}, is a rearrangement of the elements of an ordered list $\mathcal{X}$ into a \textit{one-to-one correspondence} with $\mathcal{X}$ itself. The number of permutations on a set of $n$ elements is given by $n!$ \cite{HH 12, JVU 37}. For example, there are $2!=2*1=2$ permutations of $\{1,2\}$, namely $(1,2)$ and $(2,1)$, and $3!=3*2*1=6$ permutations of $\{1,2,3\}$, namely $(1,2,3)$, $(1,3,2)$, $(2,1,3)$, $(2,3,1)$, $(3,1,2)$, and $(3,2,1)$ \cite{WWE 29}.

\subsection{Permutation methods}
	Several algorithms have been developed since, to generate all the possible permutations of $N$ elements. Most of them have been compared to see what is their best implementations on real computer, on surveys published in the field in 1960 by D.H. Lehmer \cite{DHL 60}, in 1970-71 by R.J. Ord-Smith \cite{RJOS 70, RJOS 71} and in 1977 by R. Sedgewick \cite{RS 77}. To have an overview of the history, here we show some studied methods.\\

In \cite{RS 77}, Sedgewick have review many permutation generation algorithms. From that paper, a natural way to permute an array of elements on a computer is to exchange two of its elements. The fastest permutation algorithms operate in this way: all $n!$ permutations of $n$ elements are produced by a sequence of $n! - 1$ exchanges. Several techniques using exchange such as recursive method and adjacent exchange is then described. With recursive method, for an array P, to generate all permutations of P[1],$\cdots$, P[n]. We repeat n times the step:
"First generate all permutations of P[1], $\cdots$, P[n-1], then exchange P[n] with one of the elements P[1],$\cdots$, P[n-1]". As this is repeated, a new value is put into P[n] each time. The various methods differ in their approaches to filling P[n] with the n original elements. One of the earliest algorithm based on this method was published by M.B. Wells in 1960 \cite{MBW 61}. \\

For adjacent exchanges, perhaps the most prominent permutation enumeration algorithm was formulated in 1962 by S.M. Johnson \cite{SMJ 63} and H.F. Trotter \cite{HFT 62}, apparently independently. They discovered that it was possible to generate all $n!$ permutations of $n$ elements with $n!-1$ exchanges of adjacent elements. The method is based on the natural idea that for every permutation of $n-1$ elements we can generate $n$ permutations of $n$ elements by inserting the new element into all possible positions. \\

It is the permutation generation method that determines the order of a list of permutations. In fact, there is a natural order of all permutations called lexicographic or alphabetical order \cite{RJOS 68}. In the proper sense of the word, a list of permutations is in lexicographic order if these permutations are sorted as they would appear in a dictionary. Strictly speaking, if the $n$ items going through permutations are ordered by a precedence relation $<$, then permutation $\pi_a = \left(\pi_{a1}, \pi_{a2},\cdots, \pi_{an}\right)$ precedes permutation $\pi_b = \left(\pi_{b1}, \pi_{b2},\cdots, \pi_{bn}\right)$ if and only if, for some $ i \geq 1$, we have $\pi_{aj} = \pi_{bj}$ for all $j < i$ and $\pi_{ai} < \pi_{bi}$ \cite{RMGGS 00}. Furthermore, there is a kind of "reverse lexicographic" ordering \cite{RS 77}, also called "reverse colex order" \cite{DEK 05}, which is the result of reading the lexicographic sequence backwards and the permutations from right to left.\\

Although various methods have been proposed to generate permutations in lexicographic order, they can be classified into two categories \cite{CDS 90, RMGGS 00, MKS 63, RJOS 68}. Some of these methods require the generation of the next permutation from the beginning while others produce it by a small modification of the predecessor permutation \cite{NW 78}. C.T. Djamegni and M. Tchuente \cite{DT 97} proposed in 1997 an algorithm to solve the open problem of designing a cost-optimal parallel algorithm for generating permutations of $M$ elements out of the set $\left\{0, 1,\cdots,N-1\right\}$, in lexicographic order. In 2009, Ting Kuo \cite{TK 09} have proposed a new method for generating permutations in lexicographic order using ranking an unranking functions.\\

In the case where $n$ is so large, it is difficult to generate all permutations of $n$ elements. Many authors have studied random generation permutations \cite{DT 97, GDB 1967, RD 64}. In their paper, Wendy Myrvold and Frank Ruskey \cite{WF 01} propose a ranking function for the permutations on n symbols wich assigns a unique integer in the range $\left[0,n! - 1\right] $ to each of the $n!$ permutations. Also, they propose an unranking function for which, given an integer between 0 and $n!-1$, the value of the function is the permutation having this rank. This is normally done by establishing some one-to-one correspondence between a permutation and a random number between 0 and $n!-1$. Their algorithms are presented as follows.

\subsubsection{Unranking algorithm}

First of all, let's remind that a permutation of order n is an arrangement of n symbols. We denoted by $S_n$ the set of all permutations over $\left\{0, 1, 2, \cdots,n - 1\right\}$. The array $\pi [0 \cdots n-1]$ is initialized to the identity permutation (or some other permutation) and then the following loop is executed \cite{WF 01}\\
$\phantom{salut}$ {\bf{for}} $k:= n-1, n-2, \cdots,1$ {\bf{do}} \\
$\phantom{salutsalut} swap(\pi[k],\pi[rand(k)])$ ; \\
where the call $rand(k)$ should produce a random integer in the range $0 \cdots k$.\\

 This algorithm produces a permutation selected uniformly at random from amongst all permutations in $S_n$. Let $r_{n-1}, \cdots, r_1, r_0$ be the sequence of random elements produced by the algorithm, where $0 \leq r_i \leq i$. Since there are exactly $n(n - 1)(n - 2) \cdots (2)(1) = n!$ such sequences, each different sequence must produce a different permutation. Thus we should be able to unrank if we can take an integer $r$ in the range $0 \cdots n! - 1$ and turn it into a unique sequence of values $r_{n-1},\cdots,r_1, r_0$, where $0 \leq r_i \leq i$. The details are given below \cite{WF 01}.\\
	To unrank a permutation we first initialize $\pi$ to be the identity permutation: $\pi[i] = i$, for $i= 0,1, \cdots n-1$ \cite{WF 01}.\\
	
{\bf{Procedure}} $unrank (n,r,\pi)$\\
$\phantom{salut}$ {\bf{if}} $n > 0 $ {\bf{then}} \\
$\phantom{salutsalut} swap(\pi[n-1],\pi[r$ mod $n])$ ; \\
$\phantom{salutsalut} unrank(n-1,\left\lfloor r /n\right\rfloor, \pi)$ ; \\
$\phantom{salut}$ {\bf{fi}};\\ 
$\phantom{sal}${\bf{end}};

\subsubsection{Ranking algorithm}
To rank, first compute $\pi^{-1}$. This can be done by iterating \\
$\pi^{-1}[\pi[i]] = i$, for $i=0,1,\cdots, n-1$. In the algorithm below, both $\pi$ and $\pi^{-1}$ are modified \cite{WF 01}.\\

{\bf{function}} $rank (n,\pi,\pi^{-1})$:integer\\
$\phantom{salut}$ {\bf{if}} $n = 1 $ {\bf{then}} return(0) {\bf{fi}}; \\
$\phantom{salutsalut} s:=\pi[n-1]$ ; \\
$\phantom{salutsalut} swap(\pi[n-1],\pi[\pi^{-1}[n-1]])$ ; \\
$\phantom{salutsalut} swap(\pi^{-1}[s],\pi^{-1}[n-1])$ ; \\
$\phantom{salutsalut}$return$(s+n.rank(n-1,\pi, \pi^{-1}))$ ; \\
$\phantom{sal}${\bf{end}};

\section{Our Contribution}
As presented in the related work section, the main problem that all hiding techniques face in text steganography is the low embedding capacity of secret data in text files. Thus, in the frame of designing hiding techniques using text files as cover media, our work focused on how to take advantage of permutations to generate an innocent-looking list of pseudo-randomly sorted strings and thus increase its embedding capacity. \\

In that frame, we have been interested by the work of Hioki \cite{HH 12}, who defined a permutation steganography that embeds a secret message as a number corresponding to a permutation that is represented by a tuple of cover elements. He exhibited an embedding capacity equal to $log_2(n!)$. What we noticed in his work, is the fact that his embedding capacity doesn't take into account the size of the cover elements, where as in many papers (\cite{AD 09}, \cite{MV 07}, \cite{NSMKMM 01}, \cite{KW 99}, \cite{CD 97}, \cite{CD 01}, \cite{CD 02}, \cite{PW 09}, $\cdots$) the embedding capacity is expressed by the following formula:
\textit{\begin{center}
Capacity = $\frac{number\ of\ bits\ of\ secret\ message}{number\ of\ bits\ of\ stego\ cover}$
\end{center}}
Thus by applying the above formula in the work of Hioki \cite{HH 12}, one can see that, his embedding capacity will be subject to reduction according to the cover elements used to hide a secret message.\\
Also in that method, he raised the fact that data of the cover object, subject to modification in order to embed a secret message, might be detected by an attacker trying to intercept or eliminate the message. The detection can be effective if that attacker finds some unusual arrangements of data inside the stego object.\\
  
Subsequently, in our work we have looked for and selected a particular cover media with the property that data contained inside can be modify without raising, up to a certain limit, suspicions: \textit{list}, more precisely \textit{list of strings}. This choice have been guided by the work of Desoky on list-based steganography methodology \cite{AD 09}, in which data are concealed in textual lists of itemized data.\\

Also, as we wished to make use of the power of permutations, and looked for a way to scramble data differently from the standard way (standard permutation of $n$ ordered elements) used by Hioki \cite{HH 12}, we have studied permutation algorithms. In the literature, we have been interested by the unranking and ranking permutation algorithms of Wendy et al. \cite{WF 01}, to sort a given list of strings, in a way that the queueing number of a permutation allows us to get the secret message and also to find in which order that list should be reordered.

\section{Definitions}
As below, we are making use of cover lists to hide secret message, it is important to understand the structure of a list.
We define a list as follows:
\begin{definition}
\ \\
A list $L$ is a matrix of dimensions $n*m$, where $n$ is the number of lines, $m$ the number of columns and $L[i,j]$ a string of characters (0$\leq$i$<$n, 0$\leq$j$<$m).
\end{definition}

We can have as example of lists: market list, sports betting ticket, flight board, bank account history, registration lists, $\cdots$ Further forward, to illustrate our work we take as an example list, a recording sheet of all payment transactions made by the members of a development association.

\begin{definition}
\ \\
A list $L$ with semantically low dependancy between columns is a matrix of dimensions $n*m$, where $n$ is the number of lines, $m$ the number of columns, $L[i,j]$ a string of characters $(0 \leq i < n, 0 \leq j < m)$ and there is no or a low semantic relationship between any couple of columns taken from $L$.
\end{definition}

This definition is introduced to show that in order to reduce as much as possible the factors that may raise the doubt of a cover communication by an eavesdropper, it is important that the columns taken in pairs have no semantic relationship.\\

For example, a couple of columns \textit{(name, phone number)}, or \textit{(name, address)}, shortly after swapping the information will raise suspicion if the attacker notices mistakes (lack of correspondence) between a phone number or address and certain names.

The notion of \textit{semantic relationship} is similar to the definition of \textit{Functional Dependancy} seen in relational database theory\cite{AD 93}, stated as follows: \\

\begin{definition}
\ \\
Given a relation R, a set of attributes X in R is said to functionally determine another set of attributes Y, also in R, (written $X \rightarrow Y$) if, and only if, each X value in R is associated with precisely one Y value in R; R is then said to satisfy the functional dependency $X \rightarrow Y$. 
\end{definition}

In other words, if the values for the X attributes are known (say they are x), then the values for the Y attributes corresponding to x can be determined by looking them up in any tuple of R containing x.\cite{AD 93}\\

Thus, to achieve a better embedding capacity, one may ensure that the cover list selected contains columns with no functional dependency.

\section{Approach Construction}
Here we present two different approaches, using the power of permutations, to hide a secret message. Each one of these approaches is described by a stegosystem which is a couple of algorithms to hide and recover a secret message. Note that, before beginning a cover communication, Alice and Bob can decide to encrypt or not their secret messages using a symmetric or asymmetric cipher method.
\subsection{First Approach}
In this approach, to hide a secret message in a cover list containing $n$ lines and $m$ columns $(m,n > 0)$, Alice needs to split her secret message in binary blocks of decimal values less than $n!$ and hide each block in each column, using this approach's hiding method.

\subsubsection{Hiding method}
Without loss of generality, we assume that $|s|$ (binary length of the secret message) is less or equal to $m*\lfloor log_2(n!) \rfloor - 1$, which is the maximum number of bits that can be embedded in $L$. The hiding method proceeds as follows: \\
\textbf{Pre-condition:} the cover list $L$ does not have any functional dependancy.\\
\textbf{Input:} $s$: secret message; $L$: list of $m$ columns containing $n$ lines of strings. \\
\textbf{Output:} $L$: stego-list with $s$ embedded in it\\

\textbf{Step 1}: compute the number of blocks $nb$: $nb = |s|/(\lfloor log_2(n!)\rfloor - 1)$.\\

\textbf{Step 2}: divide $s$ in $nb$ blocks of length less or equal to $(\lfloor log_2(n!)\rfloor - 1)$, and for each block $block[j]$ obtained $(0 \leq j < nb \leq m)$, add a control bit with value $1$ at the beginning, to keep, if there exists, \textit{"0"} found at first bit position of $block[j]$. Thus each block would be with a maximal length equal to $\lfloor log_2(n!)\rfloor$; \\

\textbf{Step 3}: for each block $j$ compute its decimal value $dec[j]$; \\

\textbf{Step 4}: for ($int\ j = 0; j < nb; j++$) do:\\

$\phantom{salut}$ \textbf{Step 4.1}: sort column $L[*,j]$ in the ascending order;\\

$\phantom{salut}$ \textbf{Step 4.2}: generate an array $A$, filled with integers taken between $0$ and $n-1$ as follows:
\begin{center}
\textit{
for ($int\ i = 0; i <n; i++$) do $A[i] = i$;\\
}
\end{center}

$\phantom{salut}$ \textbf{Step 4.3}: use the unranking algorithm of Wendy et al. \cite{WF 01}, specified previously, to generate a permutation corresponding to the computed value $dec[j]$ as follows:
\begin{center}
	$unrank(n, dec[j], A)$;\\
\end{center}

$\phantom{salut}$ \textbf{Step 4.4}: fill $T1$, a vector that would contain strings taken from $L[*,j]$, the following way:
\begin{center}
\textit{
for ($int\ i = 0; i < n; i++$) do $T1[i] = L[A[i], j];$\\
}
\end{center}

$\phantom{salut}$ \textbf{Step 4.5}: copy $T1$ in $L[*,j]$ as follows:
\begin{center}
\textit{
for ($int\ i = 0; i < n; i++$) do $L[i, j] = T1[i];$\\
}
\end{center}

\textbf{Step 5}: return $L$;\\

\textbf{Time Complexity}: $O(m*n*log(n))$. This time complexity depends on the sort algorithm used, and here we assume the use of the quicksort algorithm \cite{LK 07}.

\subsubsection{Recovery method}
To retrieve secret message from a stego-list $L$ encoded with the above procedure using the recovery process, that list must be recovered. Then the following procedure can be applied:\\
\textbf{Input:} $L$: stego-list;\\
\textbf{Output:} $s$: secret message\\

\textbf{Step 1}: retrieve dimensions ($m$ and $n$) of $L$ and initialize $val = 0$;\\

\textbf{Step 2}: for ($int\ j = 0; j < m; j++$) do:\\

$\phantom{salut}$ \textbf{Step 2.1}: fill $T1$, a vector that would contain strings, with strings of $L[*,j]$ in the ascending order; \\

$\phantom{salut}$ \textbf{Step 2.2}: fill $T2$, with the position indexes of each string of $L[*,j]$ as follows:
\begin{center}
\textit{
for ($int\ i = 0; i <n; i++$) do $T2[i] = GetIndexOf(L[i,j], T1)$;\\
}
\end{center}

$\phantom{salut}$ \textbf{Step 2.3}: Compute $T3$ as follows:
\begin{center}
\textit{
for ($int\ i = 0; i <n; i++$) do $T3[T2[i]] = i;$ \\
}
\end{center}

$\phantom{salut}$ \textbf{Step 2.4}: use the ranking algorithm of Wendy et al. \cite{WF 01}, specified previously, to generate the rank of the permutation, corresponding to the computed value $dec[j]$ as follows:
\begin{center}
	$dec[j] = rank(n, T2, T3)$;\\
\end{center}

$\phantom{salut}$ \textbf{Step 2.5}: convert each $dec[j]$ into its binary sequence $block[j]$, $(0 \leq j < m);$\\

\textbf{Step 3}: compute $s$ with the following process: \\

$\phantom{salut}$ $s = ""$ \\
$\phantom{salut}$ $for\ (int\ j = 0 ; j < m ; j++$) do)\{ \\
$\phantom{salutsalutsal}$  \textit{remove control bit from} $block[j]$;\\
$\phantom{salutsalutsal}$  $s = s || block[j]$\\
$\phantom{salutsal}$ \}\\

\textbf{Step 4}: return the secret message $s$;\\

Where $GetIndexOf(L[i,j], T1)$, is a function that seeks and retrieves the position index of $L[i,j]$ in $T1$, $0\leq i < n$, and is described as follows:\\
\textbf{Function} GetIndexOf(Element a, Vector T): int\\
$\phantom{salut}$ boolean bool = true;\\
$\phantom{salut}$ int $i = 0$;\\
$\phantom{salut}$ int $n = |T|$;(the number of elements contained in $T$)\\
$\phantom{salut}$ while$(bool\ and\ (i < n-1))$ do\\
$\phantom{salutsalut}$ if $(a  == T[i])$ $bool = false$;\\
$\phantom{salutsalut}$ else $i++$;\\
$\phantom{salut}$ endwhile;\\
$\phantom{salut}$ return $i$;\\
\textbf{end;}\\

\textbf{Time Complexity}: $O(m*n^2)$, as at step 2.2 $GetIndexOf$ is called $n$ times and does in the worst case, $n$ comparisons.

\subsubsection{Remark}

With this approach, blocks of secret message can be independently hidden in the cover list; meaning it's not compulsory to start by embedding the first block in the first column, then move to the next block and corresponding column. Alice can randomly treat each couple $(block[j], column[j])$, $(0 \leq j \leq m)$, before returning a stego list, thus allowing it to be implemented on a parallel computer that would help reducing it's time complexity.\\

This approach is similar to Hioki's permutation steganography \cite{HH 12}, for $m=1$. Also, the inconvenient faced by this approach is due to the fact that if lines of $L$ are sorted by an eavesdropper, the secret message would be lost; thus to reduce the risk of losing that secret message, we've developed a second approach.

\subsection{Second Approach}
In this approach, as in the first one, we assume that $|s| \leq (m-1)*(\lfloor log_2(n!) \rfloor - 1)$.\\

To hide a secret message in a cover list containing $n$ lines and $m$ columns $(m,n > 0)$, Alice needs to split her secret message in binary blocks of decimal values less than $n!$ and hide each block $block[j]$'s decimal value $dec[j]$ in each column (as seen in the previous section).\\

In this particular approach, we defined and made use of a notion of \textit{critical column}, which is a column that must not be touched by the embedding process, in order to let an attacker think that nothing is wrong with the list and also to retrieve the information, even if lines of $L$ get sorted or subject to a permutation by an attacker or anybody else.\\
Usually, it is the first column that catches the attention of a reader, because it contains information such as, line numbers, names of persons, names of products, $\cdots$; and if that column is seen unsorted for no reason, a cover communication can be implied.

\subsubsection{Hiding method}
\textbf{Pre-condition:} the cover list $L$ does not have any functional dependancy.\\
\textbf{Input:} $L$: list of strings of $n$ lines and $m$ columns. \\
\textbf{Output:} $L$: stego-list with $s$ embedded in it\\

\textbf{Step 1}: fix the critical column to the first column of $L$ $(L[*,0])$;\\

\textbf{Step 2}: compute the number of blocks $nb$: $nb = |s|/(\lfloor log_2(n!)\rfloor - 1)$. Note that the maximum value of $nb$ as a critical column has been fixed is $m-1$.\\

\textbf{Step 3}: divide $s$ in $nb$ blocks of length less than $\lfloor log_2(n!)-1 \rfloor$, and for each block $block[j]$ obtained, add a control bit with value $1$ at the beginning, to keep, if there exists, \textit{"0"} found at first bit position of $block[j]$. Thus each block would be with a maximal length equal to $\lfloor log_2(n!)\rfloor$; \\

\textbf{Step 4}: for each $block[j]$ compute its decimal value $dec[j]$; \\

\textbf{Step 5}: sort lines of $L$ in the ascending (or descending) order of elements taken from that critical column;\\

\textbf{Step 6}: for ($int\ j = 1; j <= nb; j++$) do\\
	
	$\phantom{salut}$\textbf{Step 6.1}: copy in $M$ elements of $L$ from the $j^{th}$ to the $nb^{th}$ column, as follows:\\
		\textit{
			$\phantom{salutsalut}$for ($int\ k = 0; k < n; k++$)\\
			$\phantom{salutsalutsalut}$for ($int\ l = j; l <= nb; l++$) do $M[k,l] = L[k,l]$;\\
		}
		
	$\phantom{salut}$\textbf{Step 6.2}:	sort lines of $M$ in the ascending (or descending) order of elements of its $1^{st}$ column;\\

	$\phantom{salut}$\textbf{Step 6.3}: generate an array $A$, filled with integers taken between $0$ and $n-1$ as follows:
	\begin{center}
		\textit{
			for ($int\ i = 0; i <n; i++$) do $A[i] = i$;\\
		}
	\end{center}

	$\phantom{salut}$\textbf{Step 6.4}: use the unranking algorithm of Wendy et al. \cite{WF 01}, specified previously, to generate a permutation corresponding to the computed value $dec[j]$ as follows:
	\begin{center}
		$unrank(n, dec[j-1], A)$;\\
	\end{center}

	$\phantom{salut}$\textbf{Step 6.5}: fill $T1$ with strings taken from the $1^{st}$ column of $M$, $(M[*,0])$ and fill it the following way:
\begin{center}
		\textit{
			for ($int\ i = 0; i < n; i++$) do $T1[i] =M[A[i],0];$\\
		}
	\end{center}
		
	$\phantom{salut}$\textbf{Step 6.6}: copy $T1$ in $M[*,0]$:
	\begin{center}
		\textit{
			for ($int\ i = 0; i < n; i++$) do $M[i,0] = T[i]$;\\
		}
	\end{center}

	$\phantom{salut}$\textbf{Step 6.7}: copy $M$ in $L$ as follows:\\
		\textit{
			$\phantom{salutsalut}$for ($int\ k = 0; k < n; k++$)\\
			$\phantom{salutsalutsalut}$for ($int\ l = j; l <= nb; l++$) do $L[k,l] = M[k,l]$;\\
		}
	
	\textbf{Step 7}: (optional) permute randomly lines of $L$;\\
	
	\textbf{Step 8}: return $L$;\\

\textbf{Time Complexity}: $O(m^2*n*log(n))$. This time complexity depends on the sort algorithm used at step 6.2, and here we assume the use of the quicksort algorithm \cite{LK 07}.

\subsubsection{Recovery method}
As in the previous recovery method, to retrieve secret message from a stego-list $L$ encoded with the above procedure using the recovery process, that list must be recovered. Then the following procedure can be applied:\\
\textbf{Input:} $L$: stego-list.\\
\textbf{Output:} $s$: secret message\\

\textbf{Step 1}: retrieve dimensions ($m$ and $n$) of $L$ and initialize $val = 0$;\\

\textbf{Step 2}: Sort lines of $L$ in the ascending (or descending) of elements of $L[*,0]$;\\

\textbf{Step 3}: for ($int\ j = m-1; j \geq 1; j--$) do:\\

$\phantom{salut}$ \textbf{Step 3.1}: fill $T1$ with strings of $L[*,j]$ in the ascending order; \\

$\phantom{salut}$ \textbf{Step 3.2}: fill $T2$ with position indexes of each string of $L[*,j]$ as follows:
\begin{center}
\textit{
for ($int\ i = 0; i <n; i++$) do $T2[j] = GetIndexOf(L[i,j], T1)$;\\
}
\end{center}

$\phantom{salut}$ \textbf{Step 3.3}: Compute $T3$ as follows:
\begin{center}
\textit{
for ($int\ i = 0; i <n; i++$) do $T3[T2[i]] = i$; \\
}
\end{center}

$\phantom{salut}$ \textbf{Step 3.4}: use the ranking algorithm of Wendy et al. \cite{WF 01}, specified previously, to generate the rank of the permutation, corresponding to the computed value $dec[j]$ as follows:
\begin{center}
	$dec[j] = rank(n, T2, T3)$;\\
\end{center}

$\phantom{salut}$ \textbf{Step 3.5}: sort lines of $L$ in the ascending (or descending) order of elements taken from $j^{th}$ column of $L$;\\

\textbf{Step 4}: convert each $dec[j]$ into its binary value $block[j]$, $(0 \leq j < m);$\\

\textbf{Step 5}: compute $s$ with the following process: \\

$\phantom{salut}$ $s = ""$ \\
$\phantom{salut}$ $for\ (int\ j = 0 ; j < m ; j++$) do)\{ \\
$\phantom{salutsalutsal}$  \textit{remove control bit from} $block[j]$;\\
$\phantom{salutsalutsal}$  $s = s || block[j]$\\
$\phantom{salutsal}$ \}\\

\textbf{Step 6}: return the secret message $s$;\\

Where $GetIndexOf(L[i,j], T1)$, is a function that seeks and retrieves the position index of $L[i,j]$ in $T1$, $0\leq i < n$, as described in the first approach section.\\

\textbf{Time Complexity}: $O(m*n^2)$, as it depends on the use of the quicksort algorithm \cite{LK 07}.

\subsubsection{Remark}
The major advantage of this second approach over the first, is that it resists lines sorting of L. If the lexicographical or alphabetically order of lines L, depending on the critical column, is scrambled by an eavesdropper, Bob, to retrieve the secret message, just have to sort rows in the lexicographical or alphabetical order of the critical column elements before initiating the recovery method.

\subsection{Evaluation}
As in \cite{SI 12, KCS 11}, we define bit rate or hiding capacity as the size of the hidden message relative to the size of the cover. It can be formulate as follows:
\textit{\begin{center}
Capacity = $\frac{number\ of\ bits\ of\ secret\ message}{number\ of\ bits\ of\ stego\ cover}$
\end{center}}

As the number of permutations of $n$ distinct elements is $n!$, it can be represented on $log_2(n!)$ bits. Thus, by applying the hiding method, one can see that the longest secret message that can be hidden using $L$ is of bit length $m*(log_2(n!)-1)$, for the first approach and for the second approach we have precisely $(m-1)*(log_2(n!)-1)$. Without taking in consideration the size of the cover document as in \cite{HH 12}, we've achieve a better embedding capacity (ours is $m*(log_2(n!)-1)$ and theirs is $log_2(n!)$), as they've taken a line of $m$ columns as a line of \textit{one} column, which is not our case.\\

Now, if we assume $l$ to be the average length of a string contained in $L$, then the length of the stego cover would be $(n*l)$. Thus, we can express the embedding capacity by the following expression, by taking into consideration the second approach:
\begin{proposition}\ \\
Given a secret message $s \in \{0,1\}^*$ and a cover list $L$ containing $n$ lines and $m$ columns, with semantically low dependancy between columns, where each string contained in it is with average length $l$, the embedding capacity is given by: 
\begin{center} 
$Capacity =$ {\large $\frac{(m-1)*(log_2(n!)-1)}{(n*l)}$}.
\end{center}
\end{proposition}
$\blacksquare$

By using Stirling's approximation \cite{JD 91} for factorials, which is a very powerful approximation, leading to accurate results even for small values of $n$, stated as follows:
\begin{center}
$n! \sim n^n e^{-n} \sqrt{2 \pi n}$
\end{center}
We can deduce that:
\begin{center}
$Capacity =$ {\large $\frac{(m-1)*(log_2(n^n e^{-n} \sqrt{2 \pi n})-1)}{(n*l)}$}.
\end{center}
Thus we have the following corollary:
\begin{corollary}\ \\
Given a secret message $s \in \{0,1\}^*$ and a cover list $L$ containing $n$ lines and $m$ columns, with semantically low dependancy between columns, where each string contained in it is with average length $l$, the embedding capacity is given by: 
\begin{center}
$Capacity =$ {\large $\frac{(m-1)*log_2(n*e^{-1}*\sqrt{2\pi n})}{l} + \frac{-m+1}{n*l}$}.
\end{center}
\end{corollary}
$\blacksquare$

\section{Experimental results}
We conducted some experiments on our method to analyze its performance. First of all we've applied our approaches over a data sample, then made some computation and drawn a curve showing the evolution of the embedding capacity with respect of the size of the cover list. After that we computed and put in a table de the embedding capacity for lists with 3 columns and finally we've compared our results with those found in the literature.

\subsection{Experiment 1}
In this experiment we've considered as cover list $L$ (see Table~\ref{tab:DevAss}), a text file containing all the payment operations done over a certain period of time by some members of a development association, where the first column contains names of these members, the second one, the minimal amounts of money expected for each one and the remaining columns, amounts of money each one paid per month; and as secret message the following binary sequence: $s = 01000010010011110100111001001010010011110101$ $010101110010$;.

\begin{table}[h]
\begin{center}
 \begin{tabular}{|*{5}{c|}}	
 \hline
{\scriptsize \textbf{Names}}							& {\scriptsize \textbf{Expected}} 	& {\scriptsize \textbf{January}} & {\scriptsize \textbf{February}} & {\scriptsize \textbf{March}}	\\ \hline

{\scriptsize Jean Kam}				& {\scriptsize 150000}					& {\scriptsize 20000}	& {\scriptsize 30000} & {\scriptsize 80000}			\\ \hline

{\scriptsize Sly Dolce}				& {\scriptsize 35000}					& {\scriptsize 5000}	& {\scriptsize 25000} & {\scriptsize 15000}			\\ \hline

{\scriptsize Menwick}				& {\scriptsize 70000}					& {\scriptsize 35000}	& {\scriptsize 5000} & {\scriptsize 10000}			\\ \hline

{\scriptsize Sarah Dong}				& {\scriptsize 45000}					& {\scriptsize 25000}	& {\scriptsize 10000} & {\scriptsize 5000}			\\ \hline

{\scriptsize Eddy Eko}				& {\scriptsize 50000}					& {\scriptsize 15000}	& {\scriptsize 20000} & {\scriptsize 30000}			\\ \hline

{\scriptsize Lyne Wirl}				& {\scriptsize 180000}					& {\scriptsize 100000}	& {\scriptsize 50000} & {\scriptsize 70000}			\\ \hline

{\scriptsize Jack Fack}				& {\scriptsize 200000}					& {\scriptsize 80000}	& {\scriptsize 40000} & {\scriptsize 60000}			\\ \hline

{\scriptsize Farid Al}				& {\scriptsize 450000}					& {\scriptsize 150000}	& {\scriptsize 200000} & {\scriptsize 100000}			\\ \hline

 \end{tabular}
 \end{center}
 \caption{Cover list $L$}
 \label{tab:DevAss}
\end{table}

Note that, all columns of $L$ contains distinct strings.\\

\subsubsection{First Approach}
With this approach, to hide the secret message, we made use of all the columns. We proceeded as follows:

\begin{itemize}
\item We computed the maximum number of bits that can be hidden in a column: $MaxValue = \lfloor log_2(8!) \rfloor = 15\ bits$. Then saw that a maximum number of bits $(m*MaxValue = 5*15 = 75\ bits)$ than can be hidden in $L$, is greater than the size of $s$ $(|s| = 56\ bits)$; 
\item We divided $s$ in $m$ blocks of length less than $MaxValue$, and added a control bit with value $1$ at the end of each block: $b[1] = 101000010010$, $b[2] = 101111010011$, $b[3] = 110010010100$, $b[4] = 110011110101$, $b[5] = 1010101110010$;
\item We computed for each block its decimal value: $dec[1] = 2578$, $dec[2] = 3027$, $dec[3] = 3220$, $dec[4] = 3317$, $dec[5] = 5490$;
\item For each column $j$, we've applied the $dec[j]^{th}$ permutation, allowing us to hide each block $b[j]$. We've obtained a stego list presented by the Table~\ref{tab:1stApproach};
\item The embedding capacity: $Capacity (\%) = (5*log_2(8!))/(8*(8*6)) = \textbf{20\%}$. Where the average $l$ is obtained by dividing the number of characters used \textit{(247 characters)} by the number of strings used \textit{(40 strings)}.
\end{itemize}

\begin{table}[h]
\begin{center}
 \begin{tabular}{|*{5}{c|}}	
 \hline
{\scriptsize \textbf{Names}}							& {\scriptsize \textbf{Expected}} 	& {\scriptsize \textbf{January}} & {\scriptsize \textbf{February}} & {\scriptsize \textbf{March}}	\\ \hline

{\scriptsize Jean Kam} 	        & {\scriptsize 450000}	      & {\scriptsize 80000}		 & {\scriptsize 25000} 		& {\scriptsize 60000} 		\\ \hline
{\scriptsize Menwick} 	        & {\scriptsize 50000} 	      & {\scriptsize 20000}		 & {\scriptsize 20000} 		& {\scriptsize 70000} 		\\ \hline
{\scriptsize Sarah Dong}	    & {\scriptsize 180000}	      & {\scriptsize 5000}		 & {\scriptsize 5000} 		& {\scriptsize 80000} 		\\ \hline
{\scriptsize Farid AL} 	        & {\scriptsize 45000} 	      & {\scriptsize 15000}		 & {\scriptsize 500000}		& {\scriptsize 30000} 		\\ \hline
{\scriptsize Sly Dolce} 	    & {\scriptsize 150000}	      & {\scriptsize 150000} 	 & {\scriptsize 30000} 		& {\scriptsize 10000} 		\\ \hline
{\scriptsize Lyne Wirl} 	    & {\scriptsize 200000}	      & {\scriptsize 100000} 	 & {\scriptsize 250000}		& {\scriptsize 100000} 		\\ \hline
{\scriptsize Eddy Eko} 	        & {\scriptsize 35000} 	      & {\scriptsize 25000}		 & {\scriptsize 10000} 		& {\scriptsize 5000} 		\\ \hline
{\scriptsize Jack Fack}         & {\scriptsize 70000}	      & {\scriptsize 35000}	     & {\scriptsize 40000}     	& {\scriptsize 15000}		\\ \hline

 \end{tabular}
 \end{center}
 \caption{Stego list $L$ obtained from the First Approach}
 \label{tab:1stApproach}
\end{table}

\subsubsection{Second Approach}
With this approach, to hide the secret message, we made use of all the columns except the critical one, which is the first column. We proceeded as follows:
\begin{itemize}
\item We computed the maximum number of bits that can be hidden in a column: $MaxValue = \lfloor log_2(8!) \rfloor = 15\ bits$. Then saw that a maximum number of bits $(m*MaxValue = 4*15 = 60\ bits)$ than can be hidden in $L$, is greater than the size of $s$ $(|s| = 56\ bits)$; 
\item We divided $s$ in $(m-1)$ blocks of length less than $MaxValue$, and added a control bit with value $1$ at the end of each block: $b[1] = 101000010010011$, $b[2] = 111010011100100$, $b[3] = 110100100111101$, $b[4] = 101010101110010$;
\item We computed for each block its decimal value: $dec[1] = 20627$, $dec[2] = 29924$, $dec[3] = 26941$, $dec[4] = 21874$;
\item For each column $j$, except for the critical one, we've applied the $dec[j]^{th}$ permutation, allowing us to hide each block $b[j]$. We've obtained a stego list presented by the Table~\ref{tab:2ndApproach}; The different steps are shown by tables~\ref{tab:2ndApproach1},~\ref{tab:2ndApproach2},~\ref{tab:2ndApproach3},~\ref{tab:2ndApproach4},~\ref{tab:2ndApproach5},~\ref{tab:2ndApproach6},~\ref{tab:2ndApproach7}, and~\ref{tab:2ndApproach8}.
\item The embedding capacity: $ Capacity (\%) = (4*log_2(8!))/(8*(8*6)) = \textbf{16\%}$. Where the average $l$ is obtained by dividing the number of characters used \textit{(247 characters)} by the number of strings used \textit{(40 strings)}.
\end{itemize}

\newpage
\begin{table}[h]
\begin{center}
 \begin{tabular}{|*{4}{c|}}	
 \hline
{\scriptsize \textbf{Expected}} 	& {\scriptsize \textbf{January}} & {\scriptsize \textbf{February}} & {\scriptsize \textbf{March}}	\\ \hline
{\scriptsize 35000}         & {\scriptsize 5000}         & {\scriptsize 25000}         & {\scriptsize 15000}		\\ \hline
{\scriptsize 45000}         & {\scriptsize 25000}         & {\scriptsize 10000}         & {\scriptsize 5000}		\\ \hline
{\scriptsize 50000}         & {\scriptsize 15000}         & {\scriptsize 20000}         & {\scriptsize 30000}		\\ \hline
{\scriptsize 70000}         & {\scriptsize 35000}         & {\scriptsize 5000}         & {\scriptsize 10000}		\\ \hline
{\scriptsize 150000}         & {\scriptsize 20000}         & {\scriptsize 30000}         & {\scriptsize 80000}		\\ \hline
{\scriptsize 180000}         & {\scriptsize 100000}         & {\scriptsize 50000}         & {\scriptsize 70000}		\\ \hline
{\scriptsize 200000}         & {\scriptsize 80000}         & {\scriptsize 40000}         & {\scriptsize 60000}		\\ \hline
{\scriptsize 450000}         & {\scriptsize 150000}         & {\scriptsize 200000}         & {\scriptsize 100000}		\\ \hline

 \end{tabular}
 \end{center}
 \caption{Ascending sort with respect of the element of the $1^{st}$ column}
 \label{tab:2ndApproach1}
\end{table}

\begin{table}[h]
\begin{center}
 \begin{tabular}{|*{4}{c|}}	
 \hline
{\scriptsize \textbf{Expected}} 	& {\scriptsize \textbf{January}} & {\scriptsize \textbf{February}} & {\scriptsize \textbf{March}}	\\ \hline
{\scriptsize 180000}         & {\scriptsize 5000}         & {\scriptsize 25000}         & {\scriptsize 15000}		\\ \hline
{\scriptsize 150000}         & {\scriptsize 25000}         & {\scriptsize 10000}         & {\scriptsize 5000}		\\ \hline
{\scriptsize 450000}         & {\scriptsize 15000}         & {\scriptsize 20000}         & {\scriptsize 30000}		\\ \hline
{\scriptsize 35000}         & {\scriptsize 35000}         & {\scriptsize 5000}         & {\scriptsize 10000}		\\ \hline
{\scriptsize 45000}         & {\scriptsize 20000}         & {\scriptsize 30000}         & {\scriptsize 80000}		\\ \hline
{\scriptsize 200000}         & {\scriptsize 100000}         & {\scriptsize 50000}         & {\scriptsize 70000}		\\ \hline
{\scriptsize 50000}         & {\scriptsize 80000}         & {\scriptsize 40000}         & {\scriptsize 60000}		\\ \hline
{\scriptsize 70000}         & {\scriptsize 150000}         & {\scriptsize 200000}         & {\scriptsize 100000}		\\ \hline

 \end{tabular}
 \end{center}
 \caption{Permutation of the $1^{st}$ column}
 \label{tab:2ndApproach2}
\end{table}

\newpage

\begin{table}[h]
\begin{center}
 \begin{tabular}{|*{4}{c|}}	
 \hline
{\scriptsize \textbf{Expected}} 	& {\scriptsize \textbf{January}} & {\scriptsize \textbf{February}} & {\scriptsize \textbf{March}}	\\ \hline
{\scriptsize 180000}         & {\scriptsize 5000}         & {\scriptsize 25000}         & {\scriptsize 15000}		\\ \hline
{\scriptsize 450000}         & {\scriptsize 15000}         & {\scriptsize 20000}         & {\scriptsize 30000}		\\ \hline
{\scriptsize 45000}         & {\scriptsize 20000}         & {\scriptsize 30000}         & {\scriptsize 80000}		\\ \hline
{\scriptsize 150000}         & {\scriptsize 25000}         & {\scriptsize 10000}         & {\scriptsize 5000}		\\ \hline
{\scriptsize 35000}         & {\scriptsize 35000}         & {\scriptsize 5000}         & {\scriptsize 10000}		\\ \hline
{\scriptsize 50000}         & {\scriptsize 80000}         & {\scriptsize 40000}         & {\scriptsize 60000}		\\ \hline
{\scriptsize 200000}         & {\scriptsize 100000}         & {\scriptsize 50000}         & {\scriptsize 70000}		\\ \hline
{\scriptsize 70000}         & {\scriptsize 150000}         & {\scriptsize 200000}         & {\scriptsize 100000}		\\ \hline

 \end{tabular}
 \end{center}
 \caption{Ascending sort with respect of the element of the $2^{nd}$ column}
 \label{tab:2ndApproach3}
\end{table}

\begin{table}[h]
\begin{center}
 \begin{tabular}{|*{4}{c|}}	
 \hline
{\scriptsize \textbf{Expected}} 	& {\scriptsize \textbf{January}} & {\scriptsize \textbf{February}} & {\scriptsize \textbf{March}}	\\ \hline
{\scriptsize 180000}         & {\scriptsize 80000}         & {\scriptsize 25000}         & {\scriptsize 15000}		\\ \hline
{\scriptsize 450000}         & {\scriptsize 100000}         & {\scriptsize 20000}         & {\scriptsize 30000}		\\ \hline
{\scriptsize 45000}         & {\scriptsize 25000}         & {\scriptsize 30000}         & {\scriptsize 80000}		\\ \hline
{\scriptsize 150000}         & {\scriptsize 15000}         & {\scriptsize 10000}         & {\scriptsize 5000}		\\ \hline
{\scriptsize 35000}         & {\scriptsize 150000}         & {\scriptsize 5000}         & {\scriptsize 10000}		\\ \hline
{\scriptsize 50000}         & {\scriptsize 5000}         & {\scriptsize 40000}         & {\scriptsize 60000}		\\ \hline
{\scriptsize 200000}         & {\scriptsize 20000}         & {\scriptsize 50000}         & {\scriptsize 70000}		\\ \hline
{\scriptsize 70000}         & {\scriptsize 35000}         & {\scriptsize 200000}         & {\scriptsize 100000}		\\ \hline

 \end{tabular}
 \end{center}
 \caption{Permutation of the $2^{nd}$ column}
 \label{tab:2ndApproach4}
\end{table}

\newpage

\begin{table}[h]
\begin{center}
 \begin{tabular}{|*{4}{c|}}	
 \hline
{\scriptsize \textbf{Expected}} 	& {\scriptsize \textbf{January}} & {\scriptsize \textbf{February}} & {\scriptsize \textbf{March}}	\\ \hline
{\scriptsize 35000}         & {\scriptsize 150000}         & {\scriptsize 5000}         & {\scriptsize 10000}		\\ \hline
{\scriptsize 150000}         & {\scriptsize 15000}         & {\scriptsize 10000}         & {\scriptsize 5000}		\\ \hline
{\scriptsize 450000}         & {\scriptsize 100000}         & {\scriptsize 20000}         & {\scriptsize 30000}		\\ \hline
{\scriptsize 180000}         & {\scriptsize 80000}         & {\scriptsize 25000}         & {\scriptsize 15000}		\\ \hline
{\scriptsize 45000}         & {\scriptsize 25000}         & {\scriptsize 30000}         & {\scriptsize 80000}		\\ \hline
{\scriptsize 50000}         & {\scriptsize 5000}         & {\scriptsize 40000}         & {\scriptsize 60000}		\\ \hline
{\scriptsize 200000}         & {\scriptsize 20000}         & {\scriptsize 50000}         & {\scriptsize 70000}		\\ \hline
{\scriptsize 70000}         & {\scriptsize 35000}         & {\scriptsize 200000}         & {\scriptsize 100000}		\\ \hline

 \end{tabular}
 \end{center}
 \caption{Ascending sort with respect of the element of the $3^{rd}$ column}
 \label{tab:2ndApproach5}
\end{table}

\begin{table}[h]
\begin{center}
 \begin{tabular}{|*{4}{c|}}	
 \hline
{\scriptsize \textbf{Expected}} 	& {\scriptsize \textbf{January}} & {\scriptsize \textbf{February}} & {\scriptsize \textbf{March}}	\\ \hline
{\scriptsize 35000}         & {\scriptsize 150000}         & {\scriptsize 25000}         & {\scriptsize 10000}		\\ \hline
{\scriptsize 150000}         & {\scriptsize 15000}         & {\scriptsize 20000}         & {\scriptsize 5000}		\\ \hline
{\scriptsize 450000}         & {\scriptsize 100000}         & {\scriptsize 200000}         & {\scriptsize 30000}		\\ \hline
{\scriptsize 180000}         & {\scriptsize 80000}         & {\scriptsize 30000}         & {\scriptsize 15000}		\\ \hline
{\scriptsize 45000}         & {\scriptsize 25000}         & {\scriptsize 50000}         & {\scriptsize 80000}		\\ \hline
{\scriptsize 50000}         & {\scriptsize 5000}         & {\scriptsize 10000}         & {\scriptsize 60000}		\\ \hline
{\scriptsize 200000}         & {\scriptsize 20000}         & {\scriptsize 5000}         & {\scriptsize 70000}		\\ \hline
{\scriptsize 70000}         & {\scriptsize 35000}         & {\scriptsize 40000}         & {\scriptsize 100000}		\\ \hline

 \end{tabular}
 \end{center}
 \caption{Permutation of the $3^{rd}$ column}
 \label{tab:2ndApproach6}
\end{table}

\newpage

\begin{table}[h]
\begin{center}
 \begin{tabular}{|*{4}{c|}}	
 \hline
{\scriptsize \textbf{Expected}} 	& {\scriptsize \textbf{January}} & {\scriptsize \textbf{February}} & {\scriptsize \textbf{March}}	\\ \hline
{\scriptsize 150000}         & {\scriptsize 15000}         & {\scriptsize 20000}         & {\scriptsize 5000}		\\ \hline
{\scriptsize 35000}         & {\scriptsize 150000}         & {\scriptsize 25000}         & {\scriptsize 10000}		\\ \hline
{\scriptsize 180000}         & {\scriptsize 80000}         & {\scriptsize 30000}         & {\scriptsize 15000}		\\ \hline
{\scriptsize 450000}         & {\scriptsize 100000}         & {\scriptsize 200000}         & {\scriptsize 30000}		\\ \hline
{\scriptsize 50000}         & {\scriptsize 5000}         & {\scriptsize 10000}         & {\scriptsize 60000}		\\ \hline
{\scriptsize 200000}         & {\scriptsize 20000}         & {\scriptsize 5000}         & {\scriptsize 70000}		\\ \hline
{\scriptsize 45000}         & {\scriptsize 25000}         & {\scriptsize 50000}         & {\scriptsize 80000}		\\ \hline
{\scriptsize 70000}         & {\scriptsize 35000}         & {\scriptsize 40000}         & {\scriptsize 100000}		\\ \hline

 \end{tabular}
 \end{center}
 \caption{Ascending sort with respect of the element of the $4^{th}$ column}
 \label{tab:2ndApproach7}
\end{table}

\begin{table}[h]
\begin{center}
 \begin{tabular}{|*{4}{c|}}	
 \hline
{\scriptsize \textbf{Expected}} 	& {\scriptsize \textbf{January}} & {\scriptsize \textbf{February}} & {\scriptsize \textbf{March}}	\\ \hline
{\scriptsize 150000}         	& {\scriptsize 15000}         	& {\scriptsize 20000}         	& {\scriptsize 100000}		\\ \hline
{\scriptsize 35000}         	& {\scriptsize 150000}         	& {\scriptsize 25000}         	& {\scriptsize 30000}		\\ \hline
{\scriptsize 180000}         	& {\scriptsize 80000}         	& {\scriptsize 30000}         	& {\scriptsize 80000}		\\ \hline
{\scriptsize 450000}         	& {\scriptsize 100000}         	& {\scriptsize 200000}         	& {\scriptsize 10000}		\\ \hline
{\scriptsize 50000}         	& {\scriptsize 5000}         	& {\scriptsize 10000}         	& {\scriptsize 70000}		\\ \hline
{\scriptsize 200000}         	& {\scriptsize 20000}         	& {\scriptsize 5000}         	& {\scriptsize 5000}		\\ \hline
{\scriptsize 45000}         	& {\scriptsize 25000}         	& {\scriptsize 50000}         	& {\scriptsize 60000}		\\ \hline
{\scriptsize 70000}         	& {\scriptsize 35000}         	& {\scriptsize 40000}         	& {\scriptsize 15000}		\\ \hline

 \end{tabular}
 \end{center}
 \caption{Permutation of the $4^{th}$ column}
 \label{tab:2ndApproach8}
\end{table}

\newpage

\begin{table}[h]
\begin{center}
 \begin{tabular}{|*{5}{c|}}	
 \hline
{\scriptsize \textbf{Names}}							& {\scriptsize \textbf{Expected}} 	& {\scriptsize \textbf{January}} & {\scriptsize \textbf{February}} & {\scriptsize \textbf{March}}	\\ \hline

{\scriptsize Eddy Eko}        &		     {\scriptsize 150000}         	& {\scriptsize 15000}         	& {\scriptsize 20000}         	& {\scriptsize 100000}		\\ \hline
{\scriptsize Farid AL}        &          {\scriptsize 35000}         	& {\scriptsize 150000}         	& {\scriptsize 25000}         	& {\scriptsize 30000}		\\ \hline
{\scriptsize Jack Fack}       &          {\scriptsize 180000}         	& {\scriptsize 80000}         	& {\scriptsize 30000}         	& {\scriptsize 80000}		\\ \hline
{\scriptsize Jean Kam}        &          {\scriptsize 450000}         	& {\scriptsize 100000}         	& {\scriptsize 200000}         	& {\scriptsize 10000}		\\ \hline
{\scriptsize Lyne Wirl}       &          {\scriptsize 50000}         	& {\scriptsize 5000}         	& {\scriptsize 10000}         	& {\scriptsize 70000}		\\ \hline
{\scriptsize Menwick}         &          {\scriptsize 200000}         	& {\scriptsize 20000}         	& {\scriptsize 5000}         	& {\scriptsize 5000}		\\ \hline
{\scriptsize Sarah Dong}      &          {\scriptsize 45000}         	& {\scriptsize 25000}         	& {\scriptsize 50000}         	& {\scriptsize 60000}		\\ \hline
{\scriptsize Sly Dolce}       &          {\scriptsize 70000}         	& {\scriptsize 35000}         	& {\scriptsize 40000}         	& {\scriptsize 15000}		\\ \hline

 \end{tabular}
 \end{center}
 \caption{Stego list $L$ obtained from the Second Approach}
 \label{tab:2ndApproach}
\end{table}

\subsection{Experiment 2}
\begin{figure}[h]
\centering
\includegraphics[scale=0.70]{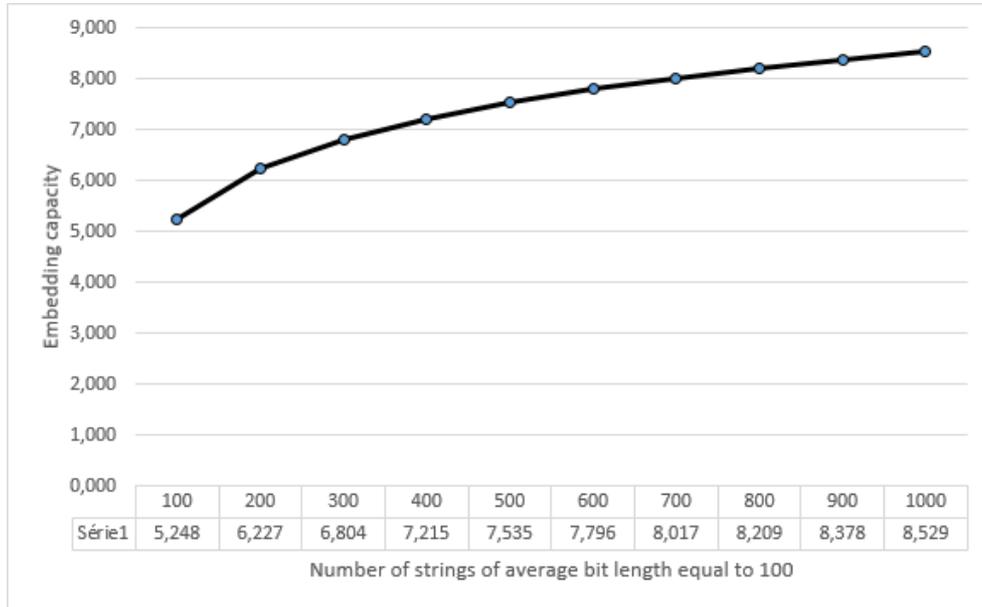} 
\caption{Evolution of the embedding capacity with respect of the size of the cover list $L$}
\label{fig:EmbeddingCapacityEvolution}
\end{figure}

By fixing the average bit length of a string to $100$ bits and varying the length of the cover list $L$ as shown on Figure~\ref{fig:EmbeddingCapacityEvolution} and we've noticed that, to increase the embedding capacity and thus attain optimal efficiency, the length of $L$ must be greater than the average bit length of strings of $L$.\\

Here, one can see that, as $l$ is fixed, the more $n$ grows away from the the average length of a string $l$, the more the embedding capacity grows. So in the condition that $n>>l$, an optimal embedding capacity can be obtained.
\newpage

\subsection{Experiment 3}
With experiment 2 we've shown that, the more the length of a list grows the more the embedding capacity does and with experiment 1 that the growth is related to the number of columns found in the cover list.\\

Obtained results exposed the fact that, if we couple the length of a list with the number of columns contained in it, we can reach a greater embedding capacity. Thus, by fixing the average bit length of a string to $100$ bits, as in experiment 2 and varying the length of the cover list $L$ containing 3 columns we've obtained the following embedding capacities, shown on Table~\ref{tab:Variation3Cols}: 

\begin{table}[h]
\begin{center}
 \begin{tabular}{|*{6}{c|}}			
 \hline
{\scriptsize \textbf{Size of $L$}} & {\scriptsize 100} & {\scriptsize 200} & {\scriptsize 300} & {\scriptsize 400} & {\scriptsize 500}\\ \hline
{\scriptsize \textbf{Embedding capacity}} & {\scriptsize $15,744\%$} & {\scriptsize $18,681\%$} & {\scriptsize $20,412\%$} & {\scriptsize $21,645\%$} & {\scriptsize $22,605\%$} \\ \hline
 \end{tabular}
  \begin{tabular}{|*{6}{c|}}			
 \hline
{\scriptsize \textbf{Size of $L$}} & {\scriptsize 600} & {\scriptsize 700} & {\scriptsize 800} & {\scriptsize 900} & {\scriptsize 1000}\\ \hline
{\scriptsize \textbf{Embedding capacity}} & {\scriptsize $23,388\%$} & {\scriptsize $24,051\%$} & {\scriptsize $24,627\%$} & {\scriptsize $25,134\%$} & {\scriptsize $25,587\%$} \\ \hline
 \end{tabular}
 \end{center}
 \caption{Variation of the embedding capacity for lists with 3 columns}
 \label{tab:Variation3Cols}
\end{table}
\ \\

Several examples can be taken, in order to increase the embedding capacity, as there exists many kind of documents in our environment, containing big lists, with numerous columns, thus allowing it. We can, for example have:
\begin{itemize}
\item list of emails, taking in consideration, the sender's name, the subject, the date of emission and the content of the message;
\item registration lists, for travels, bets or other, considering the name, date and time;
\item inventory of products of a warehouse, considering the name, quantities, manufacturer’s name, and any other useful information that can be played with;
\item list of cash deposit per day of clients of a bank or microfinance;
\item $\cdots$
\end{itemize}

Lists can also be constructed from tables of databases, containing functional dependancies, by creating transient tables with no dependancies, that can be used as cover media.\\
For instance, from a table\textit{ \textbf{Purchase}(\underline{ticket}, date, time, reference, quantity, unit price, total price)}, taken from a given database, where sales are done by some clients of a store, containing functional dependancies $(ticket, reference)\rightarrow quantity$ and $reference \rightarrow unit\ price$, we can construct a transient table (Table~\ref{tab:DBList}) containing \textit{reference}, and \textit{quantity} as columns, that can be used as cover in our hiding process.

\begin{table}[h]
\begin{center}
 \begin{tabular}{|*{2}{c|}}	
 \hline
{\scriptsize \textbf{reference}}	& {\scriptsize \textbf{quantity}}\\ \hline

{\scriptsize T-shirt}        	&		     {\scriptsize 5}	\\ \hline
{\scriptsize Jeans}        		&		     {\scriptsize 10}	\\ \hline
{\scriptsize Soda}        		&		     {\scriptsize 2}	\\ \hline
{\scriptsize Fried potatoes}    &		     {\scriptsize 1}	\\ \hline
{\scriptsize Water}        		&		     {\scriptsize 12}	\\ \hline

 \end{tabular}
 \end{center}
 \caption{Cover list obtained from a database, listing quantities of products for a month}
 \label{tab:DBList}
\end{table}

We can also construct a transient table containing for each product the quantity sold per day over a year, as presented by the Table~\ref{tab:DBList2}, that can hide a large amount of data.\\

\begin{table}[h]
\begin{center}
 \begin{tabular}{|*{6}{c|}}	
 \hline
{\scriptsize \textbf{reference}}	& {\scriptsize \textbf{Day 1}}	& {\scriptsize \textbf{Day 2}}	& {\scriptsize \textbf{Day 3}}	& {\scriptsize \textbf{$\cdots$}}	& {\scriptsize \textbf{Day 365}}\\ \hline

{\scriptsize T-shirt}        	&		     {\scriptsize 100}	&		     {\scriptsize 150}	&		     {\scriptsize 63}	&		     {\scriptsize $\cdots$}	&		     {\scriptsize 45}	\\ \hline
{\scriptsize Jeans}        		&		     {\scriptsize 50}	&		     {\scriptsize 96}	&		     {\scriptsize 7}	&		     {\scriptsize $\cdots$}	&		     {\scriptsize 33}	\\ \hline
{\scriptsize Soda}        		&		     {\scriptsize 45}	&		     {\scriptsize 35}	&		     {\scriptsize 14}	&		     {\scriptsize $\cdots$}	&		     {\scriptsize 10}	\\ \hline
{\scriptsize Fried potatoes}    &		     {\scriptsize 36}	&		     {\scriptsize 23}	&		     {\scriptsize 23}	&		     {\scriptsize $\cdots$}	&		     {\scriptsize 5}	\\ \hline
{\scriptsize Water}        		&		     {\scriptsize 85}	&		     {\scriptsize 75}	&		     {\scriptsize 78}	&		     {\scriptsize $\cdots$}	&		     {\scriptsize 20}	\\ \hline

 \end{tabular}
 \end{center}
 \caption{Cover list obtained from a database, listing quantities of products sold per day}
 \label{tab:DBList2}
\end{table}

Also, if a list of size $n$ with several columns, possesses a column which contains for instance only $m$ different elements, meaning that each element has an average of $(n/m)$ occurrences, it is possible to hide $(n/m)$ blocks of secret message in that column, where each block is hidden, one after one, using the $m$ elements allowing the computation of the permutation that hides it.
      
\newpage
\subsection{Comparison}
We have also compared our results with that of the recently developed data hiding schemes, as shown in the Table~\ref{tab:Comparison}, to present the effectiveness of our method. \\

Note that the embedding capacity can grow further, if Alice, the sender, choose the right compromise between the average bit length of strings contained in a cover list, the size and number of columns of that list.

\begin{table}[h]
\begin{center}
 \begin{tabularx}{\linewidth}{|l|l|l|}			
 \hline
{\scriptsize \textbf{Method}}							& {\scriptsize \textbf{Capacity (\%)}} 	& {\scriptsize \textbf{Explanation}}	\\ \hline
{\scriptsize Mimic functions \cite{PW 09}}				& {\scriptsize 1.27}					& {\scriptsize Using secret message at spamimc.com}				\\ \hline
{\scriptsize NICETEXT \cite{CD 97, CD 01, CD 02}}				& {\scriptsize 0.29}					& {\scriptsize Using the samples of referred articles}				\\ \hline
{\scriptsize Winstein \cite{KW 99}}						& {\scriptsize 0.5}						& {\scriptsize Based on the referred paper}		\\ \hline
{\scriptsize Murphy et al. \cite{MV 07}}						& {\scriptsize 0.30}						& {\scriptsize Reported in the referred paper}		\\ \hline
{\scriptsize Nakagawa et al. \cite{NSMKMM 01}}						& {\scriptsize 0.12}						& {\scriptsize Reported in the referred paper}		\\ \hline
{\scriptsize Stutsman et al.'s Translation based \cite{SAGG 06}}						& {\scriptsize 0.33}						& {\scriptsize Noted by authors in referred paper}		\\ \hline
{\scriptsize Topkara et al.'s Confusing \cite{TTA 07}}						& {\scriptsize 0.35}						& {\scriptsize Based on the referred paper}		\\ \hline
{\scriptsize Sun et al.'s L-R scheme \cite{SLH 04}}		& {\scriptsize 2.17}					& {\scriptsize Using the given sample in \cite{WCLL 09}}				\\ \hline
{\scriptsize Wang et al. \cite{WCLL 09}}				& {\scriptsize 3.53}					& {\scriptsize Using the given sample in \cite{WCLL 09}}		\\ \hline
{\scriptsize Listega \cite{AD 09}}						& {\scriptsize 3.87}					& {\scriptsize Based on the referred papers}				\\ \hline
{\scriptsize TEXTO \cite{MK 95}	}			& {\scriptsize 6.91}					& {\scriptsize Using sample message at eberl.net}		\\ \hline
{\scriptsize Satir and Isik \cite{SI 12}	}			& {\scriptsize 6.92}					& {\scriptsize Using example of the same article}		\\ \hline
{\scriptsize BWT+MTF+LZW coding algorithm \cite{KCS 11}}& {\scriptsize 7.03}					& {\scriptsize Using the same example of \cite{BGJX 11}}				\\ \hline
{\scriptsize \textbf{Our method}}						& {\scriptsize \textbf{25,587\%}}			& {\scriptsize Using a list of 3 columns of 1000}		\\
{\scriptsize \textbf{\ }}						& {\scriptsize \textbf{\ }}			& {\scriptsize strings with average length 100.}		\\ \hline

 \end{tabularx}
 \end{center}
 \caption{Comparison of methods}
 \label{tab:Comparison}
\end{table}


\section{Conclusion}
A novel approach of list-based steganogaphy have been proposed based on a list of sequences of characters pseudo-randomly reordered, in a way that it can embed a secret message that only the sender (Alice) and the receiver (Bob) can retrieve. Experimental results showed the feasibility of the proposed method and a comparative study showed that it performs better than some of the existing schemes in terms of embedding capacity. Further researches can be done to improve this model, and use it on other type of cover documents such as image, sound or video files.

\section{Acknowledgments}
This work was supported by {\it UMMISCO}, by {\it LIRIMA}, by {\it CETIC} and by the {\it University of Yaounde 1}.


\begin{thebibliography}{2}

\bibitem{AD 09} A. Desoky, "Listega: List-Based Steganography Methodology", \textit{International Journal of Information Security}, Springer-Verlag, Vol. 8, pp. 247-261, April 2009.

\bibitem{AD 93} Atzeny P. \& De Antonellis V., "Relational Database Theory", Benjamin-Cummings Publishing Co., Inc., Redwood City, CA, http://hdl.handle.net/11590/176707, 1993.

\bibitem{NW 78} A. Nijenhuis \& H.S. Wilf, "Combinatorial Algorithms: For Computers and Calculators", \textit{$2^{nd}$ Edition, Academic Press}, New York, NY, 1978.

\bibitem{MV 07} B. Murphy \& C. Vogel, "The syntax of concealment: reliable methods for plain text information hiding", \textit{Proceedings of the SPIE International Conference on Security, Steganography, and Watermarking of Multimedia Contents}, pp. 1503-1507, 2007.

\bibitem{CDS 90} C.D. Savage, "Generating permutations with k-differences", \textit{SIAM Journal Discrete Mathematics}, Vol. 3, No. 4, pp. 561-573, 1990.

\bibitem{DT 97} C.T. Djamegni, M. Tchuente, "A Cost-Optimal Pipeline Algorithm for Permutation Generation in Lexicographic Order", \textit{Journal of Parallel and Distributed Computing}, Vol. 44, No. 2, pp. 153-159, 1997.

\bibitem{DEK 05} D.E. Knuth, "The Art of Computer Programming", \textit{Fascicle 2: Generating all Tuples and Permutations}, Pearson Education, Upper Saddle River, NJ, Vol. 4, 2005.

\bibitem{DHL 60} D.H. Lehmer, "Teaching combinatorial tricks to a computer", \textit{in Proc. of Symposlum Appl. Math., Combinatorial Analysis}, American Mathematical Society, Providence, R.I, Vol. 10, pp. 179-193, 1960.

\bibitem{SI 12} E. Satir \& H. Isik, "A compression-based text steganography method", \textit{The Journal of Systems and Software}, Vol. 85, pp 2385–2394, 2012.

\bibitem{GDB 1967} G.D. Balbine, "Note on random permutations", \textit{Mathematics of Computation}, Vol. 21, No. 100, pp. 710-712, 1967.

\bibitem{GJS 83} G.J. Simmons, "The Prisoners' Problem and the Subliminal Channel" \textit{Advances in Cryptology - Proceedings of Crypto 83}, Session I, pp. 51-67, 1983.

\bibitem{HFT 62} H.F. Trotter, "Perm (Algorithm 115)", \textit{Communications of the ACM}, Vol. 5, No. 8, pp. 434-435, 1962.

\bibitem{HH 12} H. Hioki, "Data Embedding Methods Not Based on Content Modification", \textit{Multimedia Information Hiding Technologies and Methodologies for Controlling Data}, Chapter 7, pp. 272-294, 2012.

\bibitem{NSMKMM 01} H. Nakagawa, K. Sampei, T. Matsumoto, S. Kawaguchi, K. Makino \& I. Murase, "Text information hiding with preserved meaning—a case for Japanese documents", \textit{IPSJ Transactions 42}, Vol. 9, pp. 2339–2350, 2001.

\bibitem{JD 91} J. Dutka, "The early history of the factorial function", \textit{Archive for History of Exact Sciences}, Springer, Vol. 43, No. 3, pp. 225–249, 1991.

\bibitem{JVU 37} J.V. Uspensky, "Introduction to Mathematical Probability", \textit{New York \& London: McGraw-Hill Book Company, Inc.}, $1^{st}$ Edition,  pp. 18, 1937. 

\bibitem{RMGGS 00} K.H. Rosen, J.G. Michaels, J.L. Gross, J.W. Grossman, \& D.R. Shier, "Handbook of Discrete and Combinatorial Mathematics", \textit{CRC Press}, Boca Raton, FL, 2000.

\bibitem{MK 95} K. Maher, "TEXTO", ftp://ftp.funet.fi/pub/crypt/steganography/texto .tar.gz, 1995. 

\bibitem{KW 99} K. Winstein, "Lexical steganography through adaptive modulation of the word choice hash", {\it Secondary education at the Illinois Mathematics and Science Academy}, 1999.

\bibitem{BGJX 11} L. Bin, N. Guiqiang, L. Jianxin \& Z. Xue, "BWT-based Data Preprocessing for LZW", {\it Proceedings of the 2011 International Conference on Multimedia and Signal Processing (CMSP)}, Vol. 1, pp. 37-40, 2011.

\bibitem{LHZYC 00} L. Li, L. Huang, X. Zhao, W. Yang \& Z. Chen, "A Statistical Attack on a Kind of Word-Shift Text-Steganography", \textit{4th International Conference on Intelligent Information Hiding and Multimedia Signal Processing}, Harbin, China, pp.1503–1507, 2008.

\bibitem{LK 07} L. Khreisat, "QuickSort A Historical Perspective and Empirical Study", \textit{IJCSNS International Journal of Computer Science and Network Security}, Vol. 7, No. 12, pp. 54-65, 2007.

\bibitem{PD 08} L.Y. Por \& B. Delina, "Information Hiding: A new Approach of Text Steganography", \textit{7th WSEAS International Conference on Applied Computer and Applied Computational Science (ACACOS 'O8)}, Hangzhou, China, pp. 689-695, 2008.

\bibitem{PWC 12} L.Y. Por, K. Wong \& K.O. Chee, "UniSpaCh: a text based data hiding method using unicode space characters", \textit{Journal of Systems and Software}, http://dx.doi.org/10.1016/j.jss.2011.12.023, 2012.

\bibitem{MBW 61} M.B. Wells, "Generation of permutations by transposition", \textit{Mathematics of Computation}, Vol. 15, pp. 192-195, 1961.

\bibitem{MKS 63} M.K. Shen, "Generation of permutations in lexicographical order", \textit{Communications of the ACM}, Vol. 6, No. 9, pp.  517, 1963.

\bibitem{CD 97} M. Chapman \& G.I. Davida, "Hiding the hidden: a software system for concealing cipher text as innocuous text", \textit{The Proceedings of the International Conference on Information and Communications Security. Lecture Notes in Computer Science}, Vol. 1334, pp. 335–345, 1997.

\bibitem{CD 01} M. Chapman \& G.I. Davida, "A practical and effective approach to largescale automated linguistic steganography", \textit{Proceedings of the Information Security Conference (ISC ’01), Lecture Notes in Computer Science}, Vol. 2200, pp. 156–165, 2001.

\bibitem{CD 02} M. Chapman \& G.I. Davida, "Plausible deniability using automated linguistic steganography", \textit{International Conference on Infrastructure Security (InfraSec ’02). Lecture Notes in Computer Science}, Vol. 2437, pp. 276–287, 2002.

\bibitem{TTA 07} M. Topkara, U. Topkara \& M.J. Atallah, "Information hiding through errors: a confusing approach", \textit{Proceedings of SPIE International Conference on Security, Steganography, and Watermarking of Multimedia Contents}, San Jose, CA, USA, January 29–February 1, 2007.

\bibitem{HLA 02} N. Hopper, J. Langford \& L. von Ahn, "Provably secure steganography", \textit{Advances in Cryptology — CRYPTO 2002, Lecture Notes in Computer Science}, Springer, Vol. 2442, pp. 77-92, 2002.

\bibitem{PR 03} P. Richer, "Steganalysis: Detecting hidden information with computer forensic analysis", \textit{SANS Institute}, Version 1.4b, 2003.

\bibitem{PW 92} P. Wayner, "Mimic functions", \textit{Cryptologia XVI}, Vol. 3, pp. 193-214, 2009.

\bibitem{PW 02} P. Wayner, "Disappearing cryptography: Information hiding: Steganography \& watermarking", 2nd edition, pp. 81-128, 2002.

\bibitem{PW 09} P. Wayner, "Disappearing cryptography: Information hiding: Steganography \& watermarking", 3rd edition, 2009.

\bibitem{RD 64} R. Durstenfeld, "Random permutation", \textit{Communications of the ACM}, Vol. 7, No. 7, pp. 420, 1964.

\bibitem{KCS 11} R. Kumar, S. Chand \& S. Singh, "An Email based high capacity text steganography scheme using combinatorial compression", \textit{Fifth International Conference on Innovative Mobile and Internet Services in Ubiquitous Computing (IMIS)}, IEEE, pp. 503-508, 2011.


\bibitem{RS 77} R. Sedgewick, "Permutation generation methods", \textit{ACM Computing Surveys}, Vol. 9, No. 2, pp. 137-163, 1977.

\bibitem{SAGG 06} R. Stutsman, M. Atallah, C. Grothoff, K. Grothoff, "Lost in just the translation", \textit{Proceedings of the 2006 ACM Symposium on Applied Computing}, Dijon, France, April 23–27, pp. 338–345, 2006.

\bibitem{RJOS 68} R.J. Ord-Smith, "Generation of permutations in lexicographic order", \textit{Communications of the ACM}, Vol. 11, No. 2, pp. 117, 1968.

\bibitem{RJOS 70} R.J. Ord-Smith, "Generation of permutation sequences Part 1", \textit{Computer Journal}, Vol. 13, No. 3, pp. 152-155, 1970.

\bibitem{RJOS 71} R.J. Ord-Smith, "Generation of permutation sequences: Part 2", \textit{Computer Journal}, Vol. 14, No. 2, pp. 136-139, 1971.

\bibitem{SMJ 63} S.M. Johnson, "Generation of permutations by adjacent transposition", \textit{Mathematics of Computation}, Vol. 17, pp. 282-285, 1963.
 
\bibitem{GKDS 12} S.R. Govada, B.S. Kumar, M. Devarakonda \& M.J. Stephen, "Text Steganography with Multi level Shielding", \textit{International Journal of Computer Science Issues}, Vol. 9, Issue 4, No. 3, pp. 401-405, 2012.

\bibitem{TK 09} T. Kuo, "A New Method for Generating Permutations in Lexicographic Order", \textit{Journal of Science and Engineering Technology}, Vol. 5, No. 4, pp. 21-29, 2009.

\bibitem{WWE 29} Weisstein, W. Eric, "Permutation", \textit{MathWorld--A Wolfram Web Resource}, http://mathworld.wolfram.com/Permutation.html.

\bibitem{WF 01} W. Myrvold, F. Ruskey, "Ranking and Unranking permutations in linear time", {\it Information Processing Letters}, Vol. 79, pp. 281-284, 2001.

\bibitem{SLH 04} X. Sun, G. Luo, \& H. Huang, "Component-based digital watermarking of Chinese texts", {\it Proceedings of the 3rd international conference on Information security}, Shanghai, China, pp. 76-81, 2004.

\bibitem{WCLL 09} Z.H. Wang, C.C. Chang, C.C. Lin \& M.C. Li, "A reversible information hiding scheme using left-right and up-down Chinese character representation", {\it Journal of Systems and Software}, Vol. 82, No. 8, pp. 1362-1369, 2009.

\end{thebibliography}
\end{document}